\documentclass[acmsmall,screen]{acmart}
\AtBeginDocument{%
  }

\setcopyright{acmlicensed}
\copyrightyear{2018}
\acmYear{2018}
\acmDOI{XXXXXXX.XXXXXXX}
\acmConference['XX]{}{2025}{}

\acmISBN{}

\usepackage{booktabs}
\usepackage{graphicx}
\usepackage{textcomp}
\usepackage[most]{tcolorbox}
\usepackage{xcolor}
\usepackage{xspace}
\usepackage{url}
\usepackage{bm}
\usepackage{tabularx}
\usepackage{color}
\usepackage{colortbl}
\usepackage[switch]{lineno}
\usepackage{subcaption}
\usepackage{caption}
\usepackage{tcolorbox}
\usepackage{bbding}
\usepackage{multicol}
\usepackage{amsmath}
\usepackage{mathtools}
\usepackage[normalem]{ulem}
\usepackage{hyperref} 
\usepackage{url} 
\usepackage{threeparttable}
\usepackage{float} 
\usepackage{subcaption}

\usepackage{braket}

\usepackage{subcaption}
\usepackage{listings}
\usepackage{algorithm}
\usepackage{algpseudocode}
\usepackage{subcaption}

\newtheorem{definition}{Definition}

\definecolor{light-gray}{gray}{0.8}
\definecolor{light-red}{RGB}{255,155,155}
\definecolor{light-blue}{RGB}{0,155,255}
\definecolor{codegreen}{rgb}{0,0.6,0}
\definecolor{codegray}{rgb}{0.5,0.5,0.5}
\definecolor{codepurple}{rgb}{0.58,0,0.82}
\definecolor{backcolour}{rgb}{0.95,0.95,0.92}

\definecolor{sblue}{rgb}{0.36, 0.54, 0.66}

\definecolor{fig4green}{RGB}{151, 208, 119}
\definecolor{fig4blue}{RGB}{169, 196, 235}

\usepackage[colorinlistoftodos,prependcaption,textsize=tiny]{todonotes}

\usepackage{enumerate}
\usepackage[shortlabels]{enumitem}

\usepackage{tcolorbox}

\usepackage{multirow}
\usepackage{pbox}
\usepackage{tabularx}

\usepackage{listings}
\newcommand{\lstbg}[3][0pt]{{\fboxsep#1\colorbox{#2}{\strut #3}}}
\lstdefinestyle{mystyle}
{
    language = Python,
    basicstyle = {\ttfamily \color{main-color}},
    keywordstyle = {\color{blue}},
    keywordstyle = [2]{\color{blue}},
    keywordstyle = [3]{\color{yellow}},
    keywordstyle = [4]{\color{teal}},
    morekeywords = [3]{<<, >>},
    morekeywords = [4]{++},
    basicstyle=\ttfamily\footnotesize,
    commentstyle=\color{gray}\ttfamily,
    morecomment=[f][\lstbg{red!20}]-,
    morecomment=[f][\lstbg{green!20}]+,
    morecomment=[f][\lstbg{yellow!20}]++,
    morecomment=[f][\lstbg{yellow!20}]--,
    morecomment=[f][\textit]{@@},
    texcl=false
}

\lstdefinestyle{prompt_style}
{
    language = {},
    keywordstyle = {\color{blue}},
    keywordstyle = [2]{\color{blue}},
    keywordstyle = [3]{\color{yellow}},
    keywordstyle = [4]{\color{teal}},
    morekeywords = [3]{<<, >>},
    morekeywords = [4]{++},
    basicstyle=\ttfamily\footnotesize,
    commentstyle=\color{gray}\ttfamily,
    morecomment=[f][\lstbg{red!20}]-,
    morecomment=[f][\lstbg{green!20}]+,
    morecomment=[f][\lstbg{yellow!20}]++,
    morecomment=[f][\lstbg{yellow!20}]--,
    morecomment=[f][\textit]{@@},
    texcl=false,
    numbers=none,
    breakindent=0pt
}

\usepackage{etoolbox,environ}
\newtoggle{isArxivVersion}

\definecolor{fpbackcolor}{RGB}{242,242,242}
\definecolor{diffrem}{RGB}{202, 45, 49}
\definecolor{diffincl}{RGB}{0, 135, 90}
\definecolor{codepink}{RGB}{237, 2, 140}

\newboolean{showcomments}
\setboolean{showcomments}{true} 
\ifthenelse{\boolean{showcomments}}{
  \newcommand{\nbc}[3]{
    {\textcolor{#3}{\small{\bfseries{#1:\ }}\textit{#2}}}}
}{
  \newcommand{\nbc}[3]{}
}

\usepackage{amsfonts}

\usepackage{braket}

\usepackage{pifont}

\usepackage{xcolor}
\usepackage{soul}

\newcommand{\toolname}{AutoCodeSherpa\xspace}

\newcommand{\rev}[1]{{\color{black}{#1}\xspace}}

\newcommand{\hoare}[3]{\{#1\}\;#2\;\{#3\}}

\renewcommand{\sout}[1]{}

\begin{document}

\title{AutoCodeSherpa: Symbolic Explanations in AI Coding Agents}

\author{Sungmin Kang}
\authornote{Equal contribution, ordered alphabetically}
\email{sungmin@nus.edu.sg}
\affiliation{%
  \institution{National University of Singapore}
  \country{Singapore}
}

\author{Haifeng Ruan}
\authornotemark[1]
\email{haifeng.ruan@u.nus.edu}
\affiliation{%
  \institution{National University of Singapore}
  \country{Singapore}
}

\author{Abhik Roychoudhury}
\email{abhik@nus.edu.sg}
\affiliation{%
  \institution{National University of Singapore}
  \country{Singapore}
}

\renewcommand{\shortauthors}{YYY}

\begin{abstract}
  Large language model (LLM) agents integrate external tools with one or more LLMs to accomplish specific tasks. Agents have rapidly been adopted by developers, and they are starting to be deployed in industrial workflows, such as their use to fix static analysis issues from the widely used SonarQube static analyzer.
  However, the growing importance of agents means their actions carry greater impact and potential risk. Thus, to use them at scale, an additional layer of trust and evidence is necessary.
  This work presents \toolname, a technique that provides explanations of software issues in the form of symbolic formulae. Inspired by the reachability, infection, and propagation model of software faults, the explanations are composed of input, infection, and output conditions, collectively providing a specification of the issue. In practice, the symbolic explanation is implemented as a combination of a property-based test (PBT) and program-internal symbolic expressions. 
  Critically, this means our symbolic explanations are executable and can be automatically evaluated, unlike natural language explanations. Experiments show the generated conditions are highly accurate. For example, input conditions from \toolname had an accuracy of 85.7\%.
  This high accuracy makes symbolic explanations particularly useful in two scenarios.
  First, the explanations can be used in automated issue resolution environments to decide whether to accept or reject patches from issue resolution agents; \toolname could reject 2x as many incorrect patches as baselines did.
  Second, as agentic AI approaches continue to develop, program analysis driven explanations like ours can be provided to other LLM-based repair techniques which do not employ analysis to improve their output. In our experiments, our symbolic explanations could improve the plausible patch generation rate of the Agentless technique by 60\%.
\end{abstract}

\ccsdesc[500]{Software and its engineering~Software defect analysis}
\ccsdesc[500]{Software and its engineering~Software testing and debugging}
\ccsdesc[300]{Software and its engineering~Empirical software validation}

\keywords{LLM, Agent, Property-Based Testing}

\received{20 February 2007}
\received[revised]{12 March 2009}
\received[accepted]{5 June 2009}

\maketitle

\section{Introduction}
\label{sec:introduction}

Agents are software systems in which a large language model (LLM) autonomously invokes external tools to achieve user-specified goals. Among them, coding agents--which help developers perform coding tasks--have attracted particular attention from both industry~\cite{rondon2025passerine, github_copilot_agent} and academia~\cite{Ruan2025specrover, bouzenia2024repairagent}, due to their strong efficacy~\cite{jin2024agentsurvey}. Since their emergence, a substantial body of work on agents has rapidly been built, especially on those that resolve software issues~\cite{yang2024sweagent, Ruan2025specrover, wang2024openhands}; these agents operate on natural language issues and perform program improvements such as bug fixes or feature additions.
Industry adoption has also been rapid, with examples including Microsoft's Copilot agent ~\cite{github_copilot_agent} and AutoCodeRover~\cite{autocoderover}, which is used to fix static analysis issues in the SonarQube analyzer.

The rapid deployment of coding agents and their autonomous nature also means there is a greater potential for undesirable behavior. This makes trustworthy, evidence-based explanations for coding agents increasingly important.
\rev{Indeed, prior work demonstrates explainability and transparency are primary factors for human trust in coding agents~\cite{roychoudhury2025trustprogram}.}
The need for trustworthy explanations is also evident in real-world usage. When the Copilot agent proposed a fix for an issue in the public \texttt{.NET} repository, the lead developer was not satisfied with the patch alone and instead requested an explanation of the underlying bug, asking ``what causes us to get into this situation in the first place?''~\cite{dotnet_auto_pr}. 
\rev{More broadly}, the demand for explanations
is also strongly supported by prior literature on automated debugging.
Noller et al.~\cite{noller2022trust} find that, beyond the patch itself, the most helpful artifact an APR technique can provide is an explanation of the bug. Similarly, Kochhar et al.~\cite{Kochhar2016FLSurvey} report that over 85\% of developers consider it important that a debugging technique be able to provide its rationale.

With this need in mind, we propose \toolname, a technique that, given an issue description, generates a \emph{symbolic explanation} for why a bug occurs. Acting like an automated Sherpa\footnote{A Sherpa is a mountain guide who assists trekkers in climbing mountains.}, the symbolic explanation provides guidance through the key factors that lead to a bug.
Specifically, the explanation consists of three interrelated conditions: an \emph{input condition}, which specifies the input space under which the bug occurs; an \emph{infection condition}, which is an internal program state resulting from the bug; and an \emph{output condition}, which captures the observable symptoms of the bug. Together, these conditions help developers answer the question ``what causes us to get into this situation?''.

To generate a symbolic explanation\rev{\sout{of AI agent outputs}}, \toolname employs a multi-agent pipeline. In the first step, \toolname characterizes the input and output conditions of the bug by generating a property-based test (PBT), \rev{a generalization of failing executions}. The PBT captures the input-output properties observed in buggy executions, providing a black-box view of the circumstances under which the bug occurs. \rev{To understand the program's internal behavior,} a second agent explores the repository's code, and a third agent synthesizes infection conditions---symbolic formulae that distinguish buggy states from \rev{normal} ones. While these conditions are generated by LLMs, we \rev{apply refinement steps at each stage} to heighten accuracy.

The symbolic explanations generated by \toolname offer several unique benefits. First, as the explanation in the form of PBT is executable, it can be executed against suggested patches to
\rev{filter for likely correct ones}. This enables us to determine whether a given patch resolves the issue represented by the symbolic explanation, thereby increasing trust in the patch. In contrast, existing explainable LLM techniques~\cite{kang2025autosd, Mahbub2023Bugsplainer} produce only natural language artifacts, which cannot be executed. Second, as future software development increasingly features agent-agent interactions, on top of human-agent interactions, the explanations generated by our approach could help other agents reason about the bug and improve the likelihood of generating useful artifacts.

We
\rev{evaluate \toolname in four aspects:}
(i) \rev{the accuracy of its generated explanations,} 
(ii) automatically assessing patch correctness, (iii) assisting other coding agents in resolving bugs, \rev{and (iv) generalizing across LLMs}.
We find that the input, infection, and output condition achieve high accuracies of 85.7\%, 79.7\%, and 79.0\%, respectively, providing a basis for trust. 
Running its PBTs on agent patches, \toolname rejected 123\% more incorrect patches than the best baseline. Furthermore, the explanations were informative enough to help the issue-resolving technique Agentless~\cite{xia2024agentless}, improving its plausible patch generation rate by 60.7\%. Finally, the explanation accuracy of \toolname remained consistent across different LLMs, demonstrating the generality of the approach.

Overall, our contributions are:

\begin{itemize}
    \item \rev{Introducing executable symbolic explanations in coding agents, which describes a software issue in terms of program inputs and states symbolically and can be executed.}
    \item \toolname, an agentic tool that automatically generates symbolic explanations.
    \item Experimental evaluation showing that the symbolic explanations generated by \toolname are highly accurate, can distinguish correct from incorrect patches more accurately than existing issue reproduction techniques, and can improve the issue resolution efficacy of other techniques.
\end{itemize}


\section{Background}
\label{sec:background}

\subsection{Bug Characterization}
\label{sec:bug-characterization}
Our symbolic explanations are loosely inspired by the reachability, infection, and propagation (RIP) model of failure observation~\cite{ammann2016testbook}. The RIP model notes that for a software bug to be observable, the fault should be reached during execution (reachability), the state of the program should be incorrect (infection), and the infected state should be propagated to a statement making the state observable. This model of bugs has prominently been used in the analysis of mutation testing~\cite{just2014efficient, du2024ripples}. Our tripartite formulation of input, infection, and output conditions have a rough correspondence to the RIP model. 

Meanwhile, understanding bugs is an integral part of developers' day-to-day work and the first step
to issue resolution. Despite this importance, there is little research on associating natural language descriptions of software
issues with symbolic conditions. One naive approach to issue explanation is to prompt an LLM to generate a natural-language explanation from the issue description.
Although an explanation so generated may appear coherent and help understanding to some extent,
it is prone to errors from LLMs, even for issues in simple programs such
as introductory-level programming assignments~\cite{explanation-cs1}. To the best of our knowledge, our work is the first to produce a \emph{symbolic} explanation from natural-language issue reports.

\subsection{Hoare Triple}
\label{sec:hoare-triple}
An explanation for a bug is a description of how the bug affects the program state. A formal way of describing the propagation of program states through code is the \emph{Hoare triple}, whose standard notation is $\hoare{P}{C}{Q}$, where $P$ and $Q$ are both properties about the program state, and $C$ is a program. A Hoare triple is \emph{partially correct}, if, whenever the program $C$ starts with a state satisfying $P$ (the \emph{precondition}) \emph{and} terminates, its terminal state will satisfy $Q$ (the \emph{postcondition}). 
In contrast to partial correctness, total correctness requires reasoning that the program $C$ always terminate when the precondition $P$ is met. As deciding program termination is beyond the scope of this paper, all Hoare triples in our work are partially correct. In our work, we formalize the relationship between conditions with Hoare triples. For example, supposing the input condition is $I$, output condition is $O$, and program is $P$, we want to check that the partially correct Hoare triple $\hoare{I}{P}{O}$ holds.

\subsection{Property-Based Testing}
\label{sec:pbt}

Property-based testing (PBT) is a possible way of checking whether a Hoare triple holds. The PBT is a powerful software testing technique mainly used to find logical bugs. Beginning with the QuickCheck~\cite{quickcheck} framework for Haskell, PBT frameworks have been developed in popular programming languages like Python~\cite{hypothesis} and Java~\cite{Holser_2020}; PBTs are mature enough to have had successes in uncovering real bugs in industrial contexts~\cite{Bornholt2021,mystries-of-dropbox,testing_autosar_2015,goldstein2024property}.
In PBT, there is a \emph{property} to be checked, which is an executable specification of the program-under-test. The property often contains a \emph{precondition} that specifies the valid domain of inputs. The PBT framework automatically checks the property on a large number of randomly or semi-randomly generated inputs, which are produced by a \emph{generator} and filtered by the precondition. 
In our context, to check whether the Hoare triple $\hoare{I}{P}{O}$ holds, we use a PBT to repeatedly sample inputs that satisfy $I$, execute $P$, and check if $O$ holds.

In Figure~\ref{fig:example_pbt}, we show a simple example PBT written in the Python Hypothesis~\cite{hypothesis} framework. 
In the example, the program-under-test is a function \texttt{reciprocal} calculating the reciprocal of a floating point number. The property being tested is that a number multiplied by its reciprocal is equal to one. The precondition for the test is that the number is non-zero, and thus the generator simply produces all non-zero floating point numbers except \texttt{NaN} and infinity.

\begin{figure}
    \begin{lstlisting}
from hypothesis import given, assume
from hypothesis.strategies import floats

# PBT for a function \textasciigrave reciprocal()\textasciigrave
@given(floats(allow_nan=False, allow_infinity=False)) # generator
def test_reciprocal_property(x):
    assume(x != 0) # precondition
    # lines 9-10 are the property
    result = reciprocal(x)
    assert abs(x * result - 1.0) < 1e-9

    \end{lstlisting}
    \caption{An example PBT\label{fig:example_pbt}}
\end{figure}

PBTs stand at a useful middle ground between traditional software testing and formal methods. On one hand, because PBTs execute a large number of inputs, they are generally more rigorous than the most commonly used example-based testing, which only checks a few inputs picked by the developer. On the other hand, PBTs are usable on a wider range of programs and can be efficiently computed. PBTs have the further benefit of being syntactically similar to example-based tests, making them easy to access to developers. This is in contrast to deductive verification methods, which require specialized mathematical knowledge and thus have a higher entry barrier.

\subsection{Coding Agents}

In this work, we highlight the use of symbolic explanations for coding agents. Coding agents perform software engineering tasks by leveraging LLMs to autonomously invoke external tools. Since 2024, a variety of coding agents have emerged, differing in their levels of autonomy and their choices of tools. Representative examples include:

\begin{itemize}
\item AutoCodeRover~\cite{autocoderover}. 
AutoCodeRover follows a two-stage workflow consisting of context retrieval and patch generation. 
During context retrieval, it autonomously invokes a tool to search the program’s abstract syntax tree for relevant code snippets. 
In the patch generation stage, it produces a code change to resolve the issue. It has been integrated in the SonarQube static analyzer and deployed in production.

\item SpecRover~\cite{Ruan2025specrover}.
SpecRover is an extension of AutoCodeRover focused on patch precision: it is designed to filter out incorrect patches produced by itself. It achieves high precision by writing a reproducer test and observing its execution result, with a patch applied, in relation to the issue description. In this work, we use SpecRover as a baseline in patch validation capability due to its emphasis on precision.

\item Agentless~\cite{xia2024agentless}.
Agentless is a coding agent with minimal autonomy. 
It employs a simple three-phase process consisting of localization, repair, and patch validation. 
In each phase, the prompts to the LLM follow fixed templates. 
Due to its simplicity, Agentless is sensitive to input variations. 
Therefore, we evaluate the quality of our symbolic explanations by providing them to Agentless and observing the resulting gains in issue resolution efficacy.

\item OpenHands~\cite{wang2024openhands}.
OpenHands is a highly flexible agent capable of using a wide variety of tools to execute diverse tasks. 
For coding tasks, it primarily relies on a bash tool to execute terminal commands and a text editor tool to view and modify code. 
OpenHands has no predefined workflow and only iteratively invokes tools. As such, it has the highest level of autonomy among the coding agents introduced in this section.

\end{itemize}

While coding agents are becoming increasingly effective at resolving software issues, as demonstrated by open benchmarks~\cite{swebench-leaderboard}, the code changes they produce may still be incorrect, even when validated against a limited number of test cases~\cite{wang2025solved}. This challenge is long-standing and is known in the automated program repair (APR) community as the overfitting problem~\cite{qi2015analysis}.
To make coding agents more trustworthy, we propose symbolic explanations that provide coding agents with additional information to improve their effectiveness in resolving software issues. Moreover, these symbolic explanations can be executed to validate the code changes produced by coding agents. We evaluate both of these use cases (patch validation and improving issue resolution) in this work.

\section{Overview}
\label{sec:overview}

\begin{figure*}
    \centering
    \includegraphics[width=\linewidth]{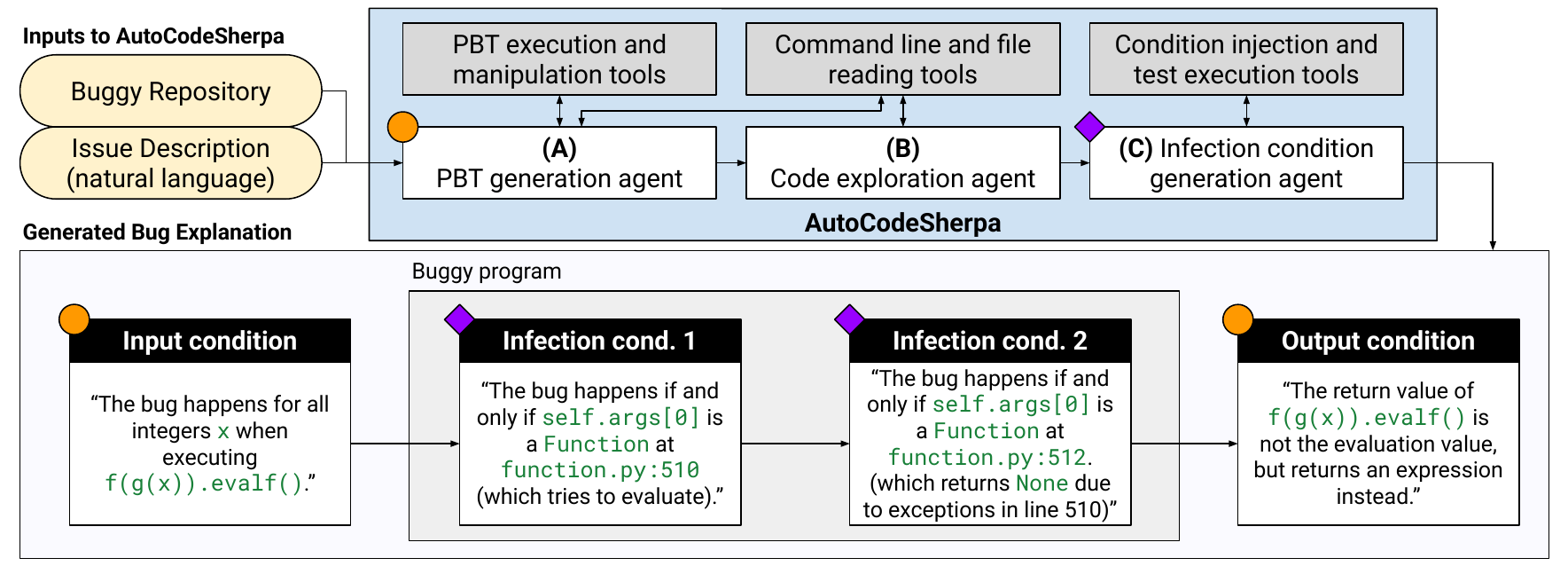}
    \caption{Overview of \toolname with a real example simplified for clarity; PBT is for property-based test.}
    \label{fig:diagram}
\end{figure*}

\lstset{
  numbers=left,
  firstnumber=500,
  numberfirstline=true
}
\begin{figure}
    \begin{lstlisting}
def _eval_evalf(self, prec):
    # Lookup mpmath function based on name
    fname = self.func.__name__
    try:
        if not hasattr(mpmath, fname):
            from sympy.utilities.lambdify import MPMATH_TRANSLATIONS
            fname = MPMATH_TRANSLATIONS[fname]
        func = getattr(mpmath, fname)
    except (AttributeError, KeyError):
        try:
            return Float(self._imp_(*self.args), prec)  # <-- BUG IS HERE
        except (AttributeError, TypeError, ValueError):
            return
...
    \end{lstlisting}
    \caption{The buggy function involved in our running example.\label{fig:running_buggy_code}}
\end{figure}

Figure~\ref{fig:diagram} presents an overview of \toolname, along with a running example used throughout this section. As illustrated in the upper part of the figure, \toolname takes as input a natural language issue description and the corresponding buggy program, and produces an explanation of the issue. Our goal is to explain why a bug occurs in a symbolic manner, inspired by the reachability--infection--propagation model described in Section~\ref{sec:bug-characterization}. To this end, \toolname employs LLM agents to identify three complementary conditions:

\begin{itemize}
\item Input condition, which specifies the set of program inputs that trigger the issue;
\item Infection condition, which describes buggy internal program states;
\item Output condition, which describes an observable fault.
\end{itemize}

\noindent \toolname also ensures that the conditions are connected by program execution: on the buggy program, execution on an input satisfying the input condition reaches a state satisfying the infection condition and then exhibits the fault specified by the output condition. This connection is checked by executing a PBT consisting of the input and output conditions while observing the values of the infection condition. A formal definition of the conditions is provided in Definition~\ref{def:symbolic_expl}.

To illustrate our approach, we present a running example highlighting the high-level operations of \toolname. The issue references a Stack Overflow post titled ``Why can't I evaluate a composition of \texttt{implemented\_functions} in SymPy at a point?''
The root cause resides in line 510 of SymPy’s \texttt{\_eval\_evalf} function, shown in Figure~\ref{fig:running_buggy_code}, while the symbolic explanation generated by \toolname is shown in Figure~\ref{fig:diagram}.

To explain the issue, \toolname first runs a PBT generation agent (Figure~\ref{fig:diagram}(A)). The agent encodes the input condition as a Python function that generates a set of inputs, and the output condition as an assertion statement. It then checks that all generated inputs trigger the assertion failure. In our example, the PBT asserts that for all integers \texttt{x} within a certain range, \texttt{f(g(x))} produces an \texttt{Expression} object instead of an expected concrete value.

Further, to explain how the bug propagates inside the program, \toolname generates infection conditions—first-order \rev{logic} formulae evaluated at specific locations to distinguish buggy states from normal ones. The code exploration agent (Figure~\ref{fig:diagram}(B)) identifies these locations by searching the code based on the issue description and the generated PBT. In our running example, the bug report already names the buggy function (Figure~\ref{fig:running_buggy_code}), allowing the agent to quickly locate it in the repository as a candidate location. We note that while this example issue has a single buggy location, the symbolic explanation technique applies to multi-location issues as well.

Within the identified buggy function, the infection condition generation agent (Figure~\ref{fig:diagram}(C)) then proceeds to produce symbolic expressions that evaluate to true only for bug-triggering inputs from the PBT. In this example, the agent generated conditions at lines 510 and 512 of the buggy function (Figure~\ref{fig:running_buggy_code}). Both conditions indicate that the bug occurs when the argument of a \texttt{Function} is itself a \texttt{Function}. They accurately capture the cause of the bug: the buggy line 510 fails to handle the argument, triggering the exception-handling code in line 512 that should not have executed.

The symbolic explanations produced by \toolname have several practical uses. First, the PBT in the symbolic explanation can be executed against candidate patches to filter out incorrect ones, a use case evaluated in Section~\ref{sec:patch-validation-eval}. 
Second, the infection conditions can enhance other software engineering techniques, such as coding agents, by generalizing beyond the issue description and revealing the internal logic of the bug. This use case is evaluated in Section~\ref{sec:efficacy-gain-eval}.


\section{Symbolic Explanations: Definition}
\label{sec:methodlogy}

%


In this section, we present the formal definition of our symbolic explanations of software issues, consisting of input, output, and infection conditions, as follows.

\begin{definition}[Symbolic explanation]
Given a program $P$ whose input parameters are $i$, an issue $X$ on the program, a symbolic
explanation for issue $X$ is a triple $(I,F_L,O)$ that satisfies the following:
\begin{itemize}
    \item Input condition $I$ is a quantifier-free first-order logic formula over $i$, and for all inputs $i$ that satisfy $I$, executing $P$ with $i$ triggers $X$;
    \item Output condition $O$ is a quantifier-free first-order logic formula over the terminal state of $P$ \rev{that holds only if $X$ is triggered}, and for which the partially correct Hoare triple $\hoare{I}{P}{O}$
    holds; thus by starting $P$ with inputs satisfying $I$, $O$ is guaranteed;
    \item Supposing $P=C_1;C_2$ where $C_1,C_2$ are sequences of program statements, and $L$ is the program location immediately after $C_1$, infection condition $F_L$ is a quantifier-free first-order logic formula over the program state at $L$, for which the partially correct Hoare triples $\hoare{I}{C_1}{F_L}$ and $\hoare{\neg I}{C_1}{\neg F_L}$ hold, i.e., $F_L$ is the result of symbolically propagating $I$ to $L$.
\end{itemize}
\label{def:symbolic_expl}
\end{definition}

By defining symbolic explanations as above, our explanations are able to provide the following information. First, input conditions describe what set of inputs trigger a software issue. Literature on bug reports shows that reproducibility, or providing inputs to trigger the issue, is a key factor in bug report quality~\cite{bachmann2009software}. Second, output conditions describe how the issue influences program output, so that readers may recognize the issue when it happens. Third, infection conditions detail how the issue propagates within the program. This helps developers, as they frequently inspect program states to better comprehend issues~\cite{alaboudi2023constitutes, beller2018dichotomy}.
Furthermore, the three conditions are connected via program execution: on the buggy program, if the input condition holds, the program must reach a state satisfying the infection condition, from which it then reaches a state satisfying the output condition.

We implement each condition as executable artifacts, making our symbolic explanations executable, unlike natural language explanations. In this work, \toolname is implemented for Python. The input and output conditions $I$ and $O$ are implemented as a PBT. Doing so allows checking whether the Hoare triple $\hoare{I}{P}{O}$ holds per Definition~\ref{def:symbolic_expl}. In particular, $I$ is implemented as a function that generates a set of inputs, and $O$ specifies an expected exception type. Infection conditions, meanwhile, are implemented as first-order logic formulae, evaluated at specific lines in Python files via runtime monitoring. The PBT framework Hypothesis~\cite{hypothesis} introduced in Section~\ref{sec:pbt} is used to repeatedly sample inputs from the input set, then check that the Hoare triples defined in Definition~\ref{def:symbolic_expl} hold.

\lstset{
  numbers=none,
  language={},
}
\begin{figure}
    \begin{lstlisting}
### Gist of the Report
The issue is that the `evalf` method in SymPy does not recursively evaluate the result of `_imp_`...

### Concrete Bug-Reproducing Inputs

1. **Input**: `f(g(2)).evalf()`
   **Expected**: `16.00000000000000`  
   **Actual**: `f(g(2))` (fails, does not evaluate recursively)

2. **Input**: `g(f(2)).evalf()`  
   **Expected**: `8.00000000000000`  
   **Actual**: `g(f(2))` (fails, does not evaluate recursively)
...
    \end{lstlisting}
    \caption{Output from the generalization phase of PBT generation.\label{fig:generalize_example}}
\end{figure}

\section{Symbolic Explanations: Construction}
\label{sec:methodology}
In this section, we present our agentic approach \toolname for generating the symbolic explanation, \rev{which involves generating a PBT for input and output conditions and program-internal expressions for infection conditions.}

\subsection{PBT Generation}
\label{sec:pbt}



\begin{figure}
    \centering
    \begin{subfigure}[b]{0.49\textwidth}
        \centering
        \includegraphics[width=\linewidth]{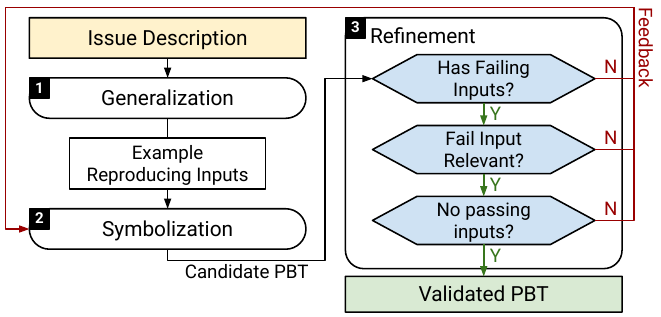}
        \caption{PBT generation agent flowchart.\label{fig:flow-pbt}}
    \end{subfigure}
    \begin{subfigure}[b]{0.49\textwidth}
        \centering
        \includegraphics[width=\linewidth]{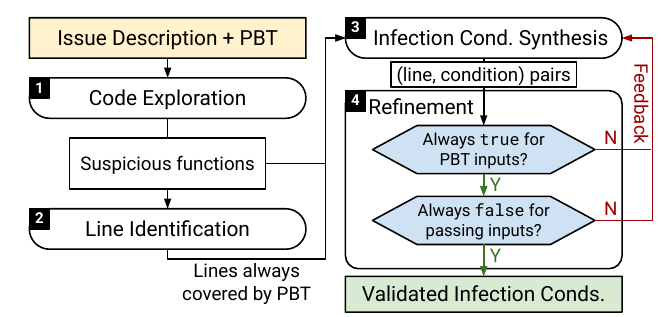}
        \caption{Infection condition generation agent flowchart.\label{fig:flow-infection}}
    \end{subfigure}
    \caption{Detailed condition generation flowcharts for \toolname.}
    \label{fig:pbt_gen_diagram}
\end{figure}

The first step in generating a symbolic explanation is to derive the input condition $I$ and output
condition $O$ from an issue description, as illustrated in Figure~\ref{fig:diagram}. This process follows a three-step agentic workflow: generalization-symbolization-refinement, as shown in Figure~\ref{fig:flow-pbt}.


\paragraph{Generalization}
To encourage generalization from the issue description, the PBT agent is first prompted to generate multiple bug-reproducing inputs based on the description, along with their corresponding actual and expected program outputs.
Figure~\ref{fig:generalize_example} illustrates the agent's response at this stage for the running example in Section~\ref{sec:overview}. While these input-output pairs may be imperfect, as they are generated by an LLM, they serve as a reasoning step from the specific input in the description to the general conditions $I$ and $O$, supporting the subsequent symbolization and refinement steps.

\paragraph{Symbolization}
With example inputs generated, the PBT agent next abstracts over them to write a PBT to reproduce the issue, using symbolic expressions to represent the input and output conditions $I$ and $O$.
As shown in Figure~\ref{fig:diagram}, the agent is equipped with tools for
command-line execution, file reading/writing, and PBT execution, enabling it
to explore relevant files and perform trial-and-error before proposing a PBT.
Once a PBT is proposed, we execute it and check for the exception specified by $O$. 
If the exception is not raised, we prompt the PBT agent to retry.  
Note that this step only ensures the \emph{existence} of an input $i$ satisfying $I$ that leads to $O$. To guarantee that \emph{all} satisfying inputs lead to $O$, i.e., that $\hoare{I}{P}{O}$ holds, the PBT is refined in the next step.

\paragraph{Refinement}
In the final step, we refine the PBT to improve its accuracy. Specifically, this step serves to guarantee that $\hoare{I}{P}{O}$ holds as per Definition~\ref{def:symbolic_expl}.
To achieve this, two types of problems need to be detected and addressed. First, some failing inputs may fail for reasons unrelated to the described issue. To handle this, we either strengthen $I$ to prevent the PBT from generating irrelevant inputs, or strengthen $O$ so that these inputs do not trigger spurious failures.
Concretely, we execute the PBT to collect failing inputs 
and their corresponding exceptions, which are then
presented to the PBT agent for review relative to the issue description. If the agent determines that any failing input $i$ is irrelevant to the described issue, the PBT is routed back to the symbolization step to generate a new PBT. Additionally, the PBT generation prompt is enriched to exclude these irrelevant failing inputs.

The second potential problem is that, even after ensuring that all failing inputs are relevant, the PBT may still generate passing inputs. To address the problem, we strengthen $I$ to filter out these inputs. 
While one could also attempt to fix the problem by weakening $O$, we avoid doing so, as this could incorrectly reject a correct program $P'$, reducing the usefulness of the symbolic explanation.
Concretely, we execute the PBT to identify passing inputs---those that satisfy $I$ but do not lead to $O$. Any passing inputs are then presented to the agent, which is prompted to revise the input condition $I$ to exclude the inputs.
Under the Hypothesis framework, the revision takes the form of either an additional \texttt{assume} statement filtering out the passing inputs or a modification to the input generator function. After revision, the PBT is executed again to check for passing inputs, and the process repeats until all inputs lead to the target failure.

\subsection{Infection Condition Generation}
\label{sec:infection}
After generating the input and output conditions, \toolname produces infection conditions $F_L$ to aid understanding of the program itself. \rev{As shown in Figure~\ref{fig:flow-infection}, infection condition generation consists of four steps, namely code exploration, line identification, infection condition synthesis, and refinement.}

\paragraph{Code Exploration}
To generate infection conditions, which are defined at specific program lines, \toolname first finds functions relevant to the bug (suspicious functions). To do so, we reuse the context retrieval agent of AutoCodeRover~\cite{autocoderover}. This agent finds likely buggy functions by invoking a tool that searches the abstract syntax tree of the program, e.g., searching for a class or for a function in a certain class. The search starts by the agent identifying keywords from the issue statement. The result of the search would reveal relevant parts of the program, and the agent would analyze the result in relation to the issue and possibly launch another search for interesting elements in the result. This process is repeated until the agent decides that the suspicious functions have been found. At this point, the identified buggy functions and the intermediate search results are passed to the infection condition agent.

\paragraph{Line Identification}

The infection condition agent receives the buggy functions identified by the code exploration agent and uses an LLM to select candidate code lines within these functions. The agent checks each candidate line to ensure it is executed by all inputs generated from the input condition $I$. If a line is not covered, the agent provides feedback to the LLM, which proposes revised lines. After up to five iterations, the agent produces a final set of lines covered by all inputs from $I$.

\paragraph{Condition Synthesis}

For each location $L$ identified in the line identification stage, the agent presents an LLM with the buggy function and the issue description, through which the LLM generates a candidate symbolic expression $F_L$. 
\rev{Specifically, the concept of infection conditions is described to the LLM in its system prompt. During operation, the LLM is then prompted to ``generate an infection condition at location [LOCATION]; you must generate general conditions that capture all bug-triggering inputs while excluding all passing tests.''
In response, the LLM generates a candidate infection condition; whether the candidate condition meets the definition of an infection condition is checked in the next step.}

\paragraph{Refinement}

A candidate infection condition can \rev{fail to meet Definition~\ref{def:symbolic_expl}} in the following two ways:
\begin{itemize}
    \item Too strong: \rev{The condition is false for at least one PBT-sampled input.} That is, there exists an input $i$ satisfying the input condition $I$ such that the infection condition $F_L$ does not hold at location $L$ for at least one visit to $L$ during execution. 
    \item Too weak: \rev{The condition is true for at least test unrelated to the bug.} That is, there exists an input $i$ that does not satisfy the input condition $I$ such that the infection condition $F_L$ holds at location $L$ for at least one visit to $L$ during execution.
\end{itemize}

\noindent
The refinement step checks whether a candidate infection condition $F_L$ is too strong or too weak.
\rev{If $F_L$ passes both checks, it is accepted. Otherwise, the agent attempts to refine it by incorporating feedback about the failure, together with the issue description and surrounding source code.}
This refinement is attempted up to five times, after which the agent moves on to the next code location.
When condition synthesis has been attempted on all code lines from the line identification stage, the agent returns the successfully identified infection conditions. 

\rev{This completes the construction of a symbolic explanation, consisting of a PBT (representing input and output conditions) and infection conditions. As a final step to achieve easier presentation, one can direct an LLM to convert the symbolic explanations to a natural language report. One can trace information in this report back to the conditions identified by \toolname, which have the properties outlined in Definition~\ref{def:symbolic_expl}, unlike natural language explanations generated by LLM prompting alone. The report can be consumed by a human developer or another coding agent.}

\section{Experimental Setup}
\label{sec:experimental_setup}


\rev{\sout{As the first work to generate symbolic explanations of issues,}} We evaluate the following research questions:

\begin{description}[leftmargin=*]\itemsep0em
    \item[RQ1:] How accurate are the the input, output and infection conditions in the explanations?

    \item[RQ2:] Are the generated PBTs useful for filtering incorrect patch candidates?

    \item[RQ3:]
    \rev{Can the symbolic explanations improve the efficacy of issue resolution techniques?
}

    \item[RQ4:] Does our symbolic explanation generation procedure generalize to different LLMs?

\sout{Among the research questions, RQ1 investigates the correctness of our symbolic explanation, which is important for the explanation to be useful;
RQ2 uses the executable nature of \toolname explanations to filter incorrect patches, while RQ3 shows the semantic value of the explanations by using them to improve another issue resolution technique. RQ4 evaluates \toolname with different LLMs to show the generalizability of \toolname across different LLMs.}

\paragraph{Benchmark}
To evaluate our approach, we use the SWE-bench Verified benchmark~\cite{chowdhury2024swebenchverified}. SWE-bench Verified is a subset of SWE-bench~\cite{jimenez2024swebench} and consists of 500 issues from 12 open source repositories that were labeled as solvable by humans.
\rev{Despite this labeling, we observe that some issues in SWE-bench Verified are under-specified, in that the expected program behavior is not fully described in the issue text.}
Each issue includes a natural-language description, which \toolname uses as input to generate explanations.

SWE-bench Verified provides two types of test cases: fail-to-pass (F2P) and pass-to-pass (P2P). F2P test cases are intended to reproduce the issue: they fail on the buggy program and should pass on the fixed version. These tests are not available to LLM agents (including \toolname) and are used only for evaluation. In contrast, P2P test cases are regression tests that pass on both the buggy and fixed programs and are available to LLM agents.


\paragraph{RQ1: Accuracy of Conditions}

In RQ1, we report how often \toolname generates a symbolic explanation for issues in the benchmark. A symbolic explanation is not produced if the LLM fails to generate an input, output, or infection condition that passes the refinement steps described in Sections~\ref{sec:pbt} and \ref{sec:infection}. For the explanations that are successfully generated, we evaluate the correctness of each condition using the following procedure:

\begin{itemize}
    \item Input and output conditions are considered correct if the PBT they form fails on the buggy program and passes on the fixed program. \rev{If this is not the case, we manually examine the input and output conditions to decide their correctness.}
    \item Infection conditions are evaluated using SWE-bench Verified test cases: an infection condition is considered correct if and only if it evaluates to true on all F2P tests and false on all P2P tests. 
\end{itemize}

\paragraph{RQ2: Patch Validation Capability}
In RQ2, we investigate whether the symbolic explanations generated by \toolname can be used to filter out incorrect patches produced by coding agents. As an example, we evaluate patches from the precision-focused agent SpecRover~\cite{Ruan2025specrover}. We report the frequency with which \toolname correctly distinguish correct patches from incorrect ones.

\paragraph{Ground-truth} To determine patch correctness, we first run the P2P and F2P test cases provided by SWE-bench Verified. Patches that pass the tests are then manually checked for semantic equivalence with the developer-written patch from the benchmark. A patch is considered correct only if it both passes the test suite and is semantically equivalent to the developer patch.

\paragraph{Baselines} We compare \toolname against two baselines: SpecRover and
OpenHands~\cite{wang2024openhands}. SpecRover predicts patch correctness based on the issue description and an agent-generated example-based reproducer test. Patches produced by SpecRover, along with their validation results, are taken from its latest public data~\cite{swebench-leaderboard}.
OpenHands is a general-purpose coding agent with state-of-the-art efficacy in reproducing issues in SWE-bench Verified~\cite{swtbench-leaderboard} among publicly available data. In particular, the best publicly available results from OpenHands were obtained using GPT-5-mini, which we also use throughout RQ1–RQ3 for a fair comparison.


\paragraph{RQ3: Efficacy Gains in Issue Resolution}
In RQ3, we evaluate an agent-agent interaction scenario, which may become more common as coding agents grow in prevalence. Specifically, we examine whether the symbolic explanations generated by \toolname improve the efficacy of Agentless in fault localization and issue resolution. For bugs where input, infection, and output conditions were all generated, an LLM converted the
symbolic explanation into a natural language report, as described in Section~\ref{sec:methodology}. This report was then supplied as additional information to Agentless, and efficacy was compared against default Agentless (without explanations) on the same set of bugs.
We choose Agentless because its simple structure makes it more sensitive to the quality of explanations than other techniques. Issue resolution efficacy is measured via the plausible patch rate, defined as the proportion of patches that pass the developer-written reproducer test.

When experimenting, we make two modifications to Agentless. First, we append the explanation to the bug report as follows: ``In addition, a trustworthy process has provided the following explanation for the bug: \{\toolname explanation\}''. Second, whereas Agentless normally generates ten patches and selects one to submit, we instead evaluate all ten patches to assess the effect of providing explanations more comprehensively.

\paragraph{RQ4: Generalization across LLMs}
In RQ1–RQ3, we reported results for \toolname generated with gpt-5-mini, the same LLM used by our best baseline, OpenHands. In RQ4, we repeat the experiments from RQ1–RQ3 using several other LLMs to assess the generalizability of \toolname across different models. In particular, we evaluate \toolname on Claude-3-5-Sonnet, a closed-weight LLM from a different provider, and DeepSeek-v3.2, an open-weight model.

\paragraph{Parameters}
The PBT agent was allowed at most 50 requests to the LLM to limit time and cost. For the code exploration agent, up to 50 invocations of the search tool were permitted, which is its default setting~\cite{specrover-repo}. The infection condition agent could suggest lines at most five times, and for each suggested line, the infection condition was iteratively improved up to five times. For fair comparison with OpenHands in RQ2, the experiments in RQ1–RQ3 were performed on gpt-5-mini.
The programs under test were set up using the official Docker images of SWE-bench Verified, with a memory limit of 6 GB per container.

\end{description}

\section{Evaluation}
\label{sec:evaluation}

This section presents the experimental results of \toolname. 

\subsection{RQ1: Accuracy of Conditions}

In RQ1, we evaluate the accuracy of the generated input, infection, and output conditions, as a necessary step in understanding the usability of the approach.

\begin{figure}
    \centering
    \includegraphics[width=\linewidth]{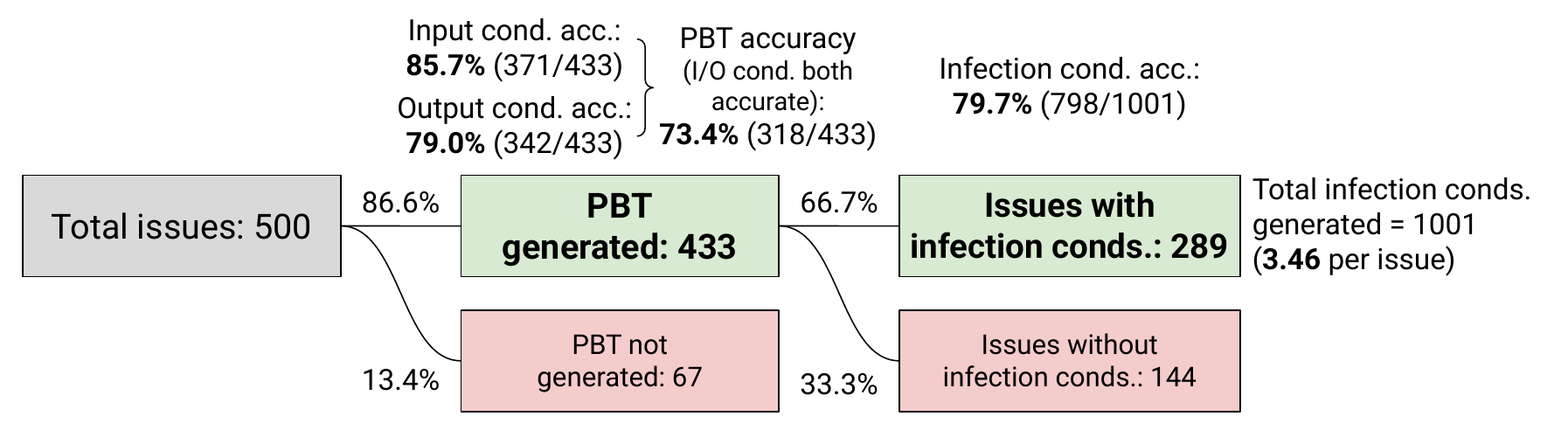}
    \caption{A plot of generation ratio and condition accuracy.}
    \label{fig:accuracy_tree}
\end{figure}

Figure~\ref{fig:accuracy_tree} presents the ratio of successful PBT generations, including both input and output conditions, alongside the accuracy of the infection conditions. We note:
\begin{itemize}
\item First, the PBTs generated by \toolname were highly accurate. A PBT specifying both the input and output conditions was generated for 433 of the 500 issues (86.6\%) in SWE-Bench Verified. In these PBTs, 85.7\% of the input conditions and 79.0\% of the output conditions were correct. In 318 (73.4\%) PBTs, both conditions were correct.
\item The infection conditions generated by \toolname were similarly accurate. Out of the 433 issues with a PBT, a total of 1001 infection conditions were generated for 289 issues (an average of 3.46 infection conditions per issue), of which 798 (79.7\%) were correct.
\item The generated symbolic explanations are thus highly accurate, and would likely be sufficient for developers to trust \toolname's results. For example, prior studies on automated debugging have shown that roughly 75\% accuracy is considered trustworthy by about 75\% of developers~\cite{Kochhar2016FLSurvey}.

\end{itemize}

Despite the high overall accuracy, we analyze the circumstances under which \toolname produces imperfect results. First, the infection condition generation rate leaves room for improvement: infection conditions were not generated for 33.3\% of all issues with generated PBTs. One challenge is the opacity of objects in certain projects. For instance, the Django project—which accounts for 46\% of the SWE-bench Verified benchmark—uses deeply nested objects, where buggy states can be subtle, making infection conditions harder to generate. Consequently, the infection condition generation rate for Django was lower compared to repositories such as matplotlib, which achieved a 90\% generation rate. Nonetheless, \toolname still generated infection conditions for a non-trivial 62\% of Django issues. Future work could improve infection condition generation by leveraging serialization libraries\footnote{https://github.com/explosion/srsly}
 to expose program state and guide condition synthesis.

\rev{Second, output conditions are the most difficult to generate accurately and have the lowest accuracy of 79.0\% among the three condition types. Some issue descriptions only cover the buggy behavior, providing hints about the input and infection conditions but leaving the intended behavior implicit or incomplete, which makes it difficult to assess the correctness of the output condition.}

\paragraph{Scalability} We further observe that \toolname scales well, yielding comparable efficacy and LLM costs across repositories of vastly different sizes. On the smallest repository in SWE-bench Verified, \texttt{requests} (30 kLoC), \toolname achieves a PBT accuracy of 81.2\%, an infection condition accuracy of 72.2\%, and an average per-issue cost of \$1.30. On the largest repository, \texttt{astropy} (804 kLoC), \toolname achieves comparable results, with a PBT accuracy of 81.8\%, an infection condition accuracy of 73.1\%, and an average per-issue cost of \$0.97.

\begin{tcolorbox}[boxrule=0pt,frame hidden,sharp corners,enhanced,borderline north={1pt}{0pt}{black},borderline south={1pt}{0pt}{black},boxsep=2pt,left=2pt,right=2pt,top=2.5pt,bottom=2pt]
    \textbf{Answer to RQ1:} \toolname is capable of generating accurate symbolic explanations for a large proportion of issues. 
\end{tcolorbox}



% ...


\subsection{RQ2: Patch Validation Capability}
\label{sec:patch-validation-eval}
\rev{\paragraph{Patch Validation Procedure} In RQ2, we evaluate the capability of our symbolic explanations to validate agent-generated patches. To do so, a patch is first applied to the buggy program, and the PBT from the explanation is executed. The underlying PBT framework repeatedly samples bug-triggering inputs and checks for test failures. If a test failure still occurs with the patch applied, the patch is considered invalid.}

\paragraph{Metrics} The symbolic explanation can be viewed as a binary classifier---patches that pass the PBT are classified as valid (positive), while those that fail are classified as invalid (negative). To assess its efficacy, we evaluate the confusion matrix of the classifier--- i.e., true positive (TP), false positive (FP), true negative (TN), and false negative (FN)---along with derived metrics such as precision and recall. As mentioned in Section~\ref{sec:experimental_setup}, evaluation is performed on patches generated for SWE-bench Verified~\cite{chowdhury2024swebenchverified} by SpecRover~\cite{Ruan2025specrover}, an LLM agent designed to improve patch precision.




\begin{table}[htb!]
\centering
\footnotesize
\begin{tabular}{rccc}
\toprule
 & \toolname & SpecRover & OpenHands \\
\midrule
\#TP $\uparrow$ & 132 & \textbf{147} & 136\\ 
\#TN $\uparrow$ & \textbf{105} & 29 & 47\\
\#FP $\downarrow$ & \textbf{140} & 216 & 198\\ 
\#FN $\downarrow$& 19 & \textbf{4} & 15 \\
\midrule
Total & \multicolumn{3}{c}{396} \\
\midrule
False positive rate\,=\,FP / (TN+FP) $\downarrow$ & \textbf{57.1}\% & 88.2\% & 80.8\% \\
Precision\,=\,TP / (TP+FP) $\uparrow$ & \textbf{48.5}\%  & 40.5\% & 40.7\% \\
Recall\,=\,TP / (TP+FN) $\uparrow$ & 87.4\%  & \textbf{97.4}\% & 90.1\% \\
\bottomrule
\end{tabular}
\caption{Confusion matrices of \toolname and baselines, where positive means correct patch.}
\label{tab:confusion-matrix}
\end{table}

\begin{table}[htb]
    \centering
    \begin{tabular}{c|c}
    \toprule
        PBT only checks for exception & {53 (36.4\%)} \\
        Issue description incomplete & {51 (36.4\%)} \\
        Developer patch goes beyond issue scope & {36 (27.2\%)} \\
    \bottomrule
    \end{tabular}
    \caption{Reasons for FP patch categorization}
    \label{tab:fp-categories}
\end{table}

Table~\ref{tab:confusion-matrix} shows the confusion matrix and related metrics for \toolname and the baselines. A total of 396 issues are included—those for which \toolname, SpecRover, and OpenHands all generated a test. We first focus on the incorrect patches (TN and FP). Of these, 245 patches (= TN + FP) are incorrect. \toolname invalidated 105 of the incorrect patches (TN), i.e., the PBT failed after their application. This number of true negatives is 262.0\% higher than SpecRover and 123.4\% higher than OpenHands, indicating that \toolname also achieves a much lower false positive rate than both baselines.

To understand the 140 false positives (FP)—i.e., incorrect patches not flagged by \toolname—we performed a manual analysis and summarized the reasons in Table~\ref{tab:fp-categories}. Among these 140 FPs, 53 PBTs only included checks for an exception, allowing any patch that resolved the exception to pass. This reflects the well-known overfitting problem in program repair, where weak test oracles can permit plausible but incorrect patches~\cite{qi2015analysis}.
For 51 issues, the issue description incompletely specifies the expected behavior—it describes only the buggy behavior but not the precise expected outputs when the bug is fixed. The PBT could not invalidate these patches because it only checks for the buggy behavior. Applying the regression tests of the program under test could further restrict patch behavior and potentially filter out 11 of these 51 incorrect patches.
Finally, for 38 issues, the patch correctly handles the reported issue (and is thus flagged as correct by the PBT), but the developer patch goes beyond the reported issue and makes additional changes to the program. In these cases, \toolname correctly allows the patches to pass according to the issue description.

Meanwhile, among the 151 (=TP+FN) correct patches, \toolname produced 19 false negatives (FN). An examination of their issue descriptions shows that 3 issues are uninformative: they provide little or no description of the buggy behavior and instead directly describe a bug fix. Due to the insufficient information, generating accurate PBTs for such issues is difficult. Another one issue describes the buggy and expected behaviors using images, whereas our agent processes text only.
For the remaining 15 FN issues, the descriptions are informative, but the generated PBTs are incorrect. These errors are caused by LLM hallucinations and do not exhibit a common pattern. Despite these FN cases, \toolname achieves a recall comparable to that of the baselines. 

Finally, we also calculate precision. Precision is important as it captures how much of developers' effort in reviewing agent-generated patches would be spent on actually correct patches. As shown in Table~\ref{tab:confusion-matrix}, \toolname has a clear margin in precision over the baselines as well. 

\begin{tcolorbox}[boxrule=0pt,frame hidden,sharp corners,enhanced,borderline north={1pt}{0pt}{black},borderline south={1pt}{0pt}{black},boxsep=2pt,left=2pt,right=2pt,top=2.5pt,bottom=2pt]
    \textbf{Answer to RQ2:}
    Compared to the SpecRover and OpenHands baselines, \toolname is more effective at identifying incorrect patches while achieving higher precision.
\end{tcolorbox}




\subsection{RQ3: Efficacy Gains in Issue Resolution}
\label{sec:efficacy-gain-eval}
In RQ3, we evaluate the efficacy gains in fault localization and issue resolution achieved by providing symbolic explanations to an example coding agent, Agentless.


\begin{table}[htb]
    \centering
    {\footnotesize
    \begin{tabular}{r|ccc}
    \toprule
                   & File top-1 & Element top-1 & Element top-any\\\midrule
Without explanations   & 65.7\% & 27.3\% & 64.4\% \\
With explanations  & \textbf{78.2\%} (+19.0\%) & \textbf{42.6\%} (+55.7\%) & \textbf{76.8\%} (+18.9\%) \\
    \bottomrule
    \end{tabular}}
    \caption{Agentless fault localization efficacy, with and without explanations}
    \label{tab:fl_improvement}
\end{table}

Table~\ref{tab:fl_improvement} shows how providing explanations from \toolname improves the fault localization accuracy of Agentless. At the file level, access to explanations increases top-1 localization accuracy by 19.0\%. For \emph{element}-level localization, where Agentless identifies suspicious classes and methods, top-1 accuracy improves by 55.7\% with explanations. For the \emph{top-any} metric, which measures whether Agentless identifies the buggy element at all, explanations enable the agent to correctly identify 18.9\% of elements that it would otherwise have missed.

Agentless also benefited from explanations when generating patches, as shown in Table~\ref{tab:bug_resolve_comparison}. As described in Section~\ref{sec:experimental_setup}, Agentless generates ten patches per issue. Considering all ten patches for all issues, access to symbolic explanations increased the number of plausible patches by 60.7\%, indicating that the explanations help guide Agentless toward effective fixes. When considering whether an issue was resolved with at least one plausible patch, Agentless successfully resolved 57.8\% of all issues, a 32.5\% improvement compared to when no explanations were provided.

\begin{table}[htb]
    \centering
    {\footnotesize
    \begin{tabular}{r|cc}
    \toprule
                   & Plausible Patch \% & Resolved Issues \\\midrule
 Without explanations   & 29.2\% & 43.6\%\\
With explanations  & \textbf{47.0\%} (+60.7\%) & \textbf{57.8}\% (+32.5\%)\\
    \bottomrule
    \end{tabular}}
    \caption{Agentless plausible patch and bug resolve rate, with and without explanations}
\label{tab:bug_resolve_comparison}
\end{table}


Despite the efficacy gains, some issues remain unresolved.
To study the reasons for failure, we investigate a sample of 30 bugs where Agentless could not generate a plausible patch despite being given an explanation. We categorize the reasons for failure as follows.
First, limitations in the benchmark itself can prevent Agentless from passing the developer test. In 30\% of the sampled cases, the test checks behavior beyond what the issue describes.
As a result, even when Agentless generates a patch consistent with the description, the developer test may still fail. For example, in \texttt{sympy\_\_sympy-21930}, the issue description mentions only one class, but the developer test also tests other classes.
For 46.7\% of sampled cases
, the explanation described how the bug occurred, but did not specify how the issue should be resolved.
Remaining minor issues included (a) Agentless 
producing patches with syntax errors (10\%); (b) the explanations being too vague (6.7\%); and (c) the explanations being misleading (6.7\%).

\begin{tcolorbox}[boxrule=0pt,frame hidden,sharp corners,enhanced,borderline north={1pt}{0pt}{black},borderline south={1pt}{0pt}{black},boxsep=2pt,left=2pt,right=2pt,top=2.5pt,bottom=2pt]
    \textbf{Answer to RQ3:} Providing Agentless with explanations from \toolname considerably improved both its fault localization and patch generation efficacy.
\end{tcolorbox}


\begin{table}[htb!]
\centering
\footnotesize
\begin{tabular}{lccc}
\toprule
\textbf{Metric} & gpt-5-mini & claude-3-5-sonnet & deepseek-v3.2 \\
\midrule
PBT Generation (RQ1) (\%)        & 86.6 & 48.4 & 57.8 \\
PBT Correctness (RQ1) (\%)       & 73.4 & 74.4 & 72.7 \\
Infection Condition Correctness (RQ1) (\%) & 79.7 & 70.3 & 85.1 \\
Patch Validation Precision (RQ2) (\%)       & 48.5 & 56.1 & 51.5 \\
Agentless Plausible Improvement (RQ3) (\%)      & +60.7 & +11.1 & +50.2 \\
\bottomrule
\end{tabular}
\caption{Evaluation of \toolname with different LLMs}
\label{tab:rq4}
\end{table}


\subsection{RQ4: Generalizability Across LLMs}
To demonstrate that \toolname generalizes to other LLMs, we ran it using different LLMs to evaluate performance, as shown in Table~\ref{tab:rq4}. \toolname could generate explanations with similar correctness, regardless of LLM: all models showed similar PBT and infection condition correctness, as well as in patch precision evaluated as in RQ2. However, the PBT generation rate differed between LLMs, indicating their capability difference. Taken together, these results suggest that even while LLMs have capability differences, the automated refinement of \toolname
(illustrated in Figure~\ref{fig:pbt_gen_diagram})
help ensure similar quality for explanations.

\begin{table}[h]
\centering
\footnotesize
\begin{tabular}{l ccc ccc ccc}
\toprule
\multirow{2}{*}{} & \multicolumn{2}{c}{First PBT} & \multicolumn{3}{c}{Refinement} & \multicolumn{2}{c}{Final PBT} \\
\cmidrule(lr){2-3} \cmidrule(lr){4-6} \cmidrule(lr){7-8}
 & Generated & Correct & Rectified & Degraded & Discarded & Generated & Correct \\
\midrule
gpt-5-mini & 477 & 246 (51.6\%) & 83 & 2 & 44 (35 incorrect) & 433 & 318 (73.4\%) \\
deepseek-v3.2 & 462 & 181 (39.2\%) & 76 & 2 & 167 (125 incorrect) &  295 & 213 (72.2\%) &  \\
\bottomrule
\end{tabular}
\caption{Effect of refinement in PBT agent. Refinement brings the correctness rate of different LLMs close. \emph{Rectified} means the first PBT was incorrect but was changed into a correct one by the refinement step, while \emph{Degraded} means a correct first PBT was changed into an incorrect one. \emph{Discarded} means a PBT was initially generated but eventually not submitted, as it could not pass the refinement step.}
\label{tab:refinement}
\end{table}

To further understand the effect of the refinement steps, we analyze the difference in correctness between the \emph{first} PBT suggested by LLMs in \toolname's execution trace, before being subjected to any \rev{refinement}, and the \emph{final} PBT which had been refined and passed the pipeline of \toolname, as shown in Table~\ref{tab:refinement}. Comparing the closed-weight LLM gpt-5-mini and the open-weight deepseek-v3.2, the first PBT correctness of the LLMs was different:
51.6\% with gpt-5-mini, and 39.2\% with deepseek-v3.2.
From this lower accuracy, \toolname provided feedback to fix initially incorrect PBTs (Rectified in Table~\ref{tab:refinement}) while preventing the submission of incorrect PBTs (Discarded). \rev{Eventually, for gpt-5-mini, 433 (=477-44) final PBTs were generated; among these, 246+83-2-(44-35)=318 were correct, where 44-35 corresponds to the number of correct first PBTs that were wrongly discarded, raising the correctness rate to 318/433 = 73.4\%. For deepseek-v3.2, although the correctness of the first PBTs were lower than that of gpt-5-mini, the refinement process discarded as many as 125 incorrect first PBTs and rectified 76. As a result, there were 462-167=295 final PBTs, among which 462+76-2-(167-125)=213 were correct, raising the correctness rate to 213/295 = 72.2\%, the same level as that of gpt-5-mini.} 

\rev{Meanwhile, all three models improved the plausible patch generation rate of Agentless. deepseek-v3.2 notably showed a strong improvement of 50.2\%, similar to gpt-5-mini. However, claude-3-5-sonnet showed lower improvement than the other two models. This is due to two factors. First, claude-3-5-sonnet produced less accurate infection conditions than the other models as in Tab.~\ref{tab:rq4}, increasing the likelihood of explanations misguiding Agentless. Second, when converting symbolic explanations from \toolname into natural language, claude-3-5-sonnet produced the shortest and vaguest explanations: its explanations were half the length of gpt-5-mini on average.}

\begin{tcolorbox}[boxrule=0pt,frame hidden,sharp corners,enhanced,borderline north={1pt}{0pt}{black},borderline south={1pt}{0pt}{black},boxsep=2pt,left=2pt,right=2pt,top=2.5pt,bottom=2pt]
    \textbf{Answer to RQ4} \toolname generalizes to different LLMs, generating symbolic explanations for all LLMs we evaluate. Furthermore, the refinement steps of \toolname ensure similar quality despite differences in LLM capability.
\end{tcolorbox}

\begin{figure*}[h!!]
    \centering
    \includegraphics[width=0.9\linewidth]{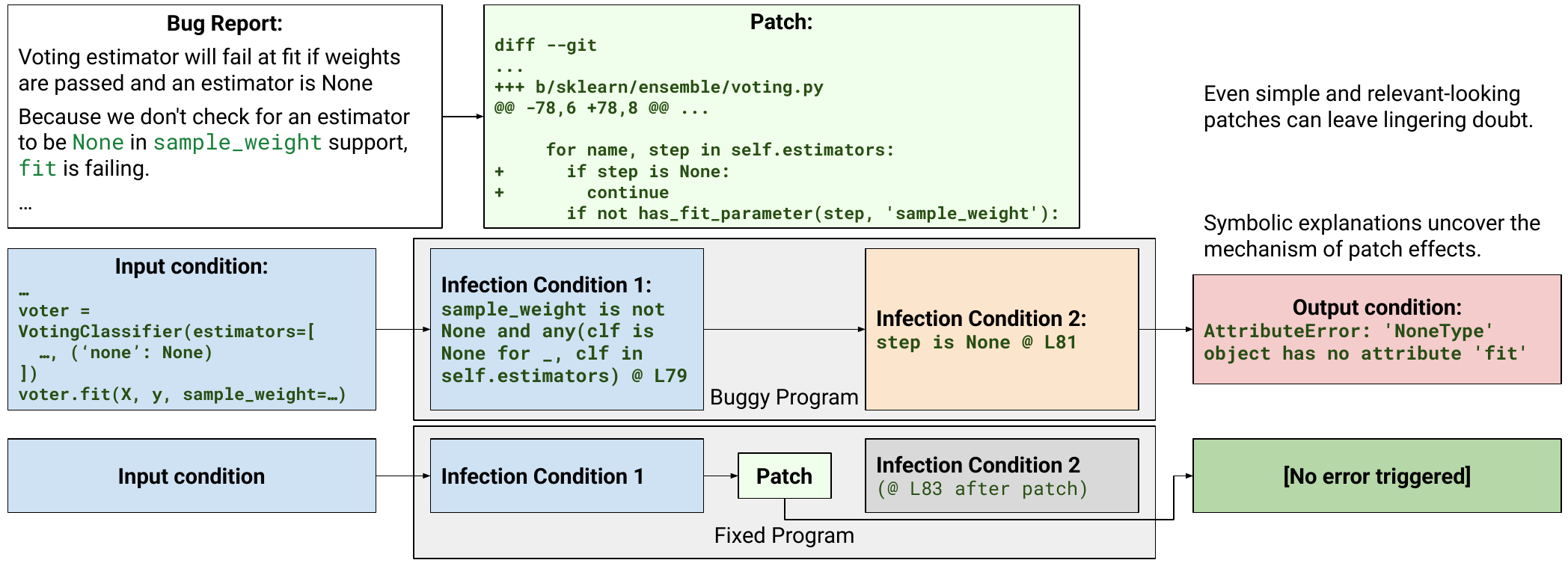}
    \caption{A schematic of a correct and helpful explanation.}
    \label{fig:good_diagram}
\end{figure*}

\section{Case Studies}
\label{sec:examples}

To illustrate the symbolic explanations generated by \toolname, we discuss two examples, one helpful and one unhelpful, and analyze them in the context of their corresponding issues.

\paragraph{Helpful Example} Figure~\ref{fig:good_diagram} presents a schematic of the issue \texttt{scikit-learn-13779} from SWE-Bench Verified and its symbolic explanation. The issue description, shown at the top-left, reports failure when the \texttt{estimator} parameter is \texttt{None}. A candidate patch, generated by SpecRover and displayed at the top, appears to handle the parameter correctly. However, in a real-world context, it would be difficult to immediately trust this patch, since LLMs are known to produce plausible but incorrect patches~\cite{sarkar2022like}.
Developers would likely need to inspect the code carefully to understand the patch’s logic and its effects.

On the bottom half of Figure~\ref{fig:good_diagram}, we illustrate how the symbolic explanation from \toolname can help understand both the issue and the patch.
First, the input condition, shown on the left, accurately indicates that the issue arises when the \texttt{estimator} parameter is \texttt{None}.
Second, infection condition 1, shown in the middle, further helps to understand the issue by indicating the values of key local variables in a buggy execution.
Third, infection condition 2 is particularly helpful for understanding the patch by SpecRover: this condition indicates that the bug is triggered when the variable \texttt{step} is \texttt{None}, which is exactly the case handled by the patch.
Finally, the PBT in the explanation can be executed to check automatically that the issue is resolved by the patch.

\paragraph{Unhelpful Example} We present an unhelpful symbolic explanation generated for the issue \texttt{django-14539}.  The issue description mentions that the \texttt{urlize} function has a bug. There are in fact two \texttt{urlize} functions in the \texttt{django} repository, and the bug report does not clarify which. In fact, only one of these two (defined in \texttt{html.py}) has real functionality; the second function (defined in \texttt{defaultfilters.py}) is simply calling the first. However, both the input condition and the generated infection condition erroneously focus on the \texttt{defaultfilters.py} file. As a result, the symbolic explanation does not provide useful information about the root cause of the bug (and the error propagation chain) within the \texttt{html.py} file. This shows that our explanations can sometimes focus on correlated phenomena, rather than conducting an accurate analysis of the causal chain responsible for the bug. The example gives a source of improvement of our work.

\section{Threats to Validity}
\label{sec:threats}

While experimental results demonstrate the effectiveness of \toolname, we note the limitations of this study. Regarding threats to internal validity, in this study we performed manual analysis to assess the correctness of patches, which may contain mistakes. To mitigate this risk, two authors of the paper independently assessed the correctness of plausible patches, comparing them with the developer patch, and discussed to make a final decision. Meanwhile, there are threats to external validity - the results presented in this work may not generalize to other programming languages or bugs outside of SWE-bench Verified, and the Agentless performance improvement results may not generalize to other issue resolution techniques. Despite this, we note that the concept of bug characterization in the form of symbolic expressions, as well as its instantiation in the form of property-based tests and within-program first-order predicates, is general.

\section{Related Work}
\label{sec:relwork}

\subsection{Issue Reproduction}





One line of work closely related to ours is issue reproduction, which aims to write
a test case to reproduce a given issue. Since our symbolic explanation
is executable, it is akin to a reproducer test. However, unlike existing
works on issue reproduction~\cite{libro,wang2024aegis,swtbench}, which only try to find example-based reproducing tests,
our explanation concisely represents a large, potentially infinite number of buggy program
executions with symbolic expressions. In other words, our symbolic explanation
is unique in that it \emph{generalizes} from the issue description.

\subsection{Automatic Bug Explanation}

Research on developer perception of automated debugging techniques has revealed that developers are \rev{sceptical} of patches provided without evidence, and that they seek explanations for the root cause of a bug~\cite{Kochhar2016FLSurvey, noller2022trust}. This observation has prompted researchers to propose techniques that explain bugs. Bugsplainer~\cite{Mahbub2023Bugsplainer} uses a neural machine translation technique to generate natural language explanations for specific software lines. AutoSD~\cite{kang2025autosd} uses LLMs to follow the scientific process to generate and validate hypotheses about the bug. SpecRover~\cite{Ruan2025specrover} analyzes the intended behavior for software components and generates a natural language explanation based on these findings. In contrast to all such approaches, \toolname generates a \emph{symbolic} bug explanation, which is executable and thus for which the veracity can be automatically evaluated.

\subsection{Input Conditions}
We are unaware of any techniques that jointly generate input, infection, and output conditions, in the manner same as ours.
However, there have been efforts to characterize the input space. For example, Avicenna~\cite{eberlein2023semantic} seeks to identify and explain the inputs that cause a fault. However, it can only be applied to inputs conforming to a predefined grammar. This is unlike our approach which uses the expressiveness of PBTs to represent general input and output conditions. As most tasks in our benchmark involve the construction and use of complex objects, we assess that Avicenna would be difficult to compare against in this work. Daikon~\cite{ernst1999daikon} generates invariants for program states, but requires these invariants to fit pre-defined property templates.
AutoSD~\cite{kang2025autosd}, a program repair technique, generates hypotheses about a bug and inspects the internal state. However, it generates explanations specific to patches from the tool, unlike the patch-agnostic and formal bug explanations from \toolname. 


\subsection{Coding Agents}
As mentioned earlier, an LLM agent is an autonomous software system that allows LLMs to invoke tools to interact with its environment. Of particular relevance to our work are coding agents, LLM agents that perform software engineering (SE) tasks. Since early examples such as SWE-Agent~\cite{yang2024sweagent} and AutoCodeRover~\cite{autocoderover}, which focus on resolving software issues, the capability of SE agents has expanded to a wide range of tasks, including test generation~\cite{swtbench}, bug finding~\cite{zheng2025llm}, and many more \cite{use-agent}. To tackle these tasks, SE agents typically make use of program analysis tools such as code search~\cite{autocoderover}, code edit, test execution, and command-line execution~\cite{yang2024sweagent}. Some agentic systems can employ multiple agents to collaborate on a complex task, with each agent specializing in one part or one step in the task~\cite{Ruan2025specrover}. In the present paper, we have proposed \toolname, a multi-agent system for the task of producing bug explanations.

\section{Perspectives}

\label{sec:conclusion}

Agents represent a promising new paradigm for the execution of software engineering (SE) activities. Armed with an LLM back-end, agents invoke various file navigation and program analysis tools to autonomously carry out SE tasks. Despite their capabilities, however, human oversight and trust-building are still required for successful deployment of SE agents. To that end, in this work we show the promise of symbolic explanations of issue reports. By producing these symbolic explanations we can enhance developer understanding, improve the quality of output produced by other agents, and most importantly produce evidence of correctness for code and patches generated by agents. Our explanations are executable so they can filter out incorrect patches produced by agents, to improve the quality of automatically generated code.  As SE agents become more commonplace, and coding witnesses further automation, our work can be seen as a mechanism to enhance trust in auto-coding. In this sense, it contributes to the vision of {\em trusted automatic programming} \rev{and its integration into future workflows.}


\section*{Acknowledgments}
This work was partially supported by a Singapore Ministry of Education (MoE) Tier 3 grant "Automated Program Repair", MOE-MOET32021-0001.

\bibliographystyle{ACM-Reference-Format}
\bibliography{references}

\end{document}